\documentclass[printer]{aa}
\usepackage{graphics,epsfig,lscape}
\usepackage{natbib,aas_macros}
\bibpunct{(}{)}{;}{a}{}{,}

\def\cm-2{cm$^{-2}$}

\def\ein{{\it Einstein}}
\def\chandra{{\it Chandra}}

\def\xmm{{XMM-Newton}}
\def\n253{\object{NGC~253}}
\def\rxj{\object{RX~J004717.4-251811}}

\newcommand{\ergs}[1]{$\times10^{#1}$ \hbox{erg s$^{-1}$}}
\newcommand{\oergs}[1]{$10^{#1}$ erg s$^{-1}$}
\newcommand{\hcm}[1]{$\times10^{#1}$ cm$^{-2}$}

\newcommand{\expo}[1]{$\times10^{#1}$}
\newcommand{\oexpo}[1]{$10^{#1}$}

\newcommand{\nh}{\hbox{$N_{\rm H}$}}

\begin{document}

   \title{RX J004717.4-251811: The first eclipsing X-ray binary outside 
    the Local Group\thanks{This work is based 
    on observations obtained with XMM-Newton, an ESA Science Mission 
    with instruments and contributions directly funded by ESA Member
    States and the USA (NASA).}}

   \author{W. Pietsch \and  
           F. Haberl \and
	   A. Vogler
          }
\institute{Max-Planck-Institut f\"ur extraterrestrische Physik, 85741 Garching, Germany 
    }
     
     \offprints{W.~Pietsch}
     \mail{wnp@mpe.mpg.de}

   \date{Received; accepted }
   \titlerunning{Eclipsing X-ray binary in \n253}

        \abstract{The X-ray source \rxj\ in the nearby starburst galaxy \n253\ 
	was found to undergo changes from a low to a high 
	state twice, during an \xmm\ EPIC observation in December 2000 and also during a 
	\chandra\ observation one year earlier. These transitions are interpreted as
	egresses from eclipses of a compact object in a high mass X-ray binary system (HMXB). 
	The phase of eclipse egress
	during the \chandra\ observation is given by barycenter corrected 
	MJD 51539.276$\pm$0.006 and the binary period determined 
	to p~=~($352.870\pm0.012$)~d~/~n by the time
	difference between the two egresses and number n of periods in-between. Allowed periods may be further 
	constrained by additional \xmm, \chandra, ROSAT, and \ein\ observations
	resulting in only seven acceptable periods with 1.47024~d and
	3.20793~d most promising. No significant regular pulsations of the 
	source in the range 0.3--1000 s were found. 
	Fluctuations on time scales of 1000 s were observed 
	together with extended intervals of low intensity. The energy spectrum during the bright state can best be described
	by an absorbed flat power law (\nh=1.9\hcm{21}, $\Gamma$=1.7). In the
	bright state the source luminosity is 4\ergs{38} (0.5--5 keV), just
	compatible with the Eddington luminosity of a 1.4 M$_{\sun}$\ 
	neutron star. A possible
	optical identification is suggested. \rxj\ parameters are compared to
	other eclipsing X-ray binaries (XRBs). 
} 
\maketitle

\section{Introduction}
Luminous X-ray binaries (XRBs) in the Milky Way were among the first X-ray 
sources detected and optically identified. They are classified as low 
mass X-ray binary (LMXBs) and high mass X-ray binary (HMXBs) systems according
to the mass of the donor star. The accreting collapsed object can be a neutron
star or a black hole in either class.

XRBs may show orbital variability (eclipses, absorption dips, regular
outbursts), pulsations due to rotation of the compact object (neutron star), 
long term variability (which may be explained due to precession of an accretion 
disk or free precession of a neutron star) and flares. 
Examples for eclipsing XRBs are Her X-1, LMC X-4, SMC X-1 with pulsation periods
of 1.24 s, 0.71 s, 13.5 s,
orbital periods of 1.7 d, 1.4 d, 3.9 d, full X-ray eclipse durations of at least 
0.137, 0.144, 0.147 in phase, long term
periods of 35 d, 30 d, 45 d -- 60 d, and  showing no flares, flares, no flares, 
respectively \citep[][ and references therein]{2000A&AS..147...25L}. 
Eclipsing XRBs up to now were only detected within the 
Local Group and the total number remains small. 
The first examples of such systems
\citep[Cen X-3 and Her X-1:][ respectively]{1972ApJ...172L..79S,1972ApJ...174L.143T}, 
were detected in the Milky Way with the UHURU satellite, 
then in the SMC \citep[SMC X-1:][]{1972ApJ...178L..71S}
and LMC \citep[LMC X-4:][]{1978Natur.271...37L,1978Natur.271...38W}, and 
finally one in M33 \citep{1989ApJ...336..140P}. Combining X-ray pulse arrival 
time information and orbital parameters of pulsating eclipsing XRBs with 
radial-velocities of the optical companion were essential to determine neutron 
star masses \citep[see e.g.][]{1995xrb..book...58V}. 

In this paper we report on the first detection of an eclipsing XRB outside the 
Local Group in \n253.   
This prototypical nearby edge-on starburst galaxy  
was observed by many X-ray satellites. 
X-ray point sources could be resolved with the high spatial resolution
instruments aboard \ein, ROSAT, \chandra, and \xmm. 
\citet[][ hereafter FT84]{1984ApJ...286..491F}
reported complex X-ray emission from eight point sources as well as 
diffuse emission from nuclear area, plume and disk using the \ein\ HRI 
detector. 
\citet[][ hereafter VP99]{1999A&A...342..101V} analyzed X-ray point 
sources in the bulge, disk and halo of \n253 in ROSAT PSPC and HRI 
observations. The number of point-like X-ray sources increased to 30 sources, 13
of which are time variable. \rxj\ (E2 of FT84, X17 of VP99) is one of the four
\ein\ sources detected again with ROSAT. During most of the ROSAT observations 
a second source of similar brightness (X18) was visible just 15\arcsec\ to the 
south. \rxj\ shows strong time variability within the ROSAT observations and 
also compared to the \ein\ HRI detection. This time 
variability and the ROSAT PSPC hardness ratio classification made \rxj\ a 
good XRB candidate. 

\n253 was observed by \xmm\ as a performance verification target in June 2000.
First results on bright point sources were discussed in 
\citet[][ hereafter P01]{2001A&A...365L.174P}. Of special interest was the spectrum and time
variability of the ROSAT source X33 30\arcsec\ south of the nucleus, most likely
a black hole XRB. The pair of ROSAT sources
(X17/X18) is clearly visible in Fig.~1 of P01, however deferred for later 
analysis, as these sources at the time could not be distinguished by the
SAS source detection algorithms. 

\n253 was again observed by us with \xmm\ in December 2000 within the \xmm\ 
survey scientist guaranteed time program. We noticed that during this 
observation the count rate of \rxj\ abruptly changed from very low to high. This
behavior most likely represents an egress from an eclipse by the companion
star in this XRB. Triggered by this finding we analyzed archival 
\chandra, \xmm, ROSAT, and \ein\ observations. This led to the detection of a
similar behavior during \chandra\ observation 790. Data from all the satellites
were then used to further constrain the binary orbit parameters and long term
time variability of the source. In the following we report on these findings.

\section{Observations}
\begin{table*}
\caption[]{\xmm, \chandra, ROSAT, and \ein\ observatory observations of \rxj.
\xmm\ count rates, hardness ratios (HR) and luminosities are determined for
the EPIC PN camera.}
\begin{tabular}{lllrrrrrr}
\hline\noalign{\smallskip}
\multicolumn{1}{c}{Observatory} &
\multicolumn{1}{c}{Obs. id.} & \multicolumn{1}{c}{Obs. dates} &
\multicolumn{1}{c}{L.t.} & \multicolumn{1}{c}{R$_{e}$} & 
\multicolumn{1}{c}{Count rate} & \multicolumn{1}{c}{HR} &
\multicolumn{1}{c}{L$_{\rm X}^{**}$} & \multicolumn{1}{c}{Comment} \\ 
\noalign{\smallskip}
& & & (ks) & (\arcsec)& (ct ks$^{-1}$) & & (\oergs{37})\\
\noalign{\smallskip}
\multicolumn{1}{c}{(1)} & \multicolumn{1}{c}{(2)} & \multicolumn{1}{c}{(3)} & 
\multicolumn{1}{c}{(4)} & \multicolumn{1}{c}{(5)} & \multicolumn{1}{c}{(6)} & 
\multicolumn{1}{c}{(7)} & \multicolumn{1}{c}{(8)} & \multicolumn{1}{c}{(9)} \\
\noalign{\smallskip}\hline\noalign{\smallskip}
\xmm     & 0125960101 & 2000-06-03 & 39.0 & 10.0 & $7.0\pm0.4$ & $0.51\pm0.05^{\dagger}$ & $2.6\pm0.2$ & \\
\xmm     & 0125960201 & 2000-06-04 & 14.0 & 10.0 & $7.2\pm0.7$ & $0.38\pm0.05^{\dagger}$ & $2.7\pm0.3$ & \\
\xmm     & 0110900101 & 2000-12-13/14 & 30.7 & 10.0 & $4.3\pm0.7$ & $0.50\pm0.15$ &  $1.4\pm0.2$ & low state\\
         &          &               &      &    & $41.8\pm1.7$ & $0.86\pm0.07$ &  $14.0\pm0.6$ & high state\\ 
\chandra & 969 & 1999-12-16 & 14.2 & 3.5 & $54.8\pm2.0$ & $0.74\pm0.06$ & $20.1\pm0.7$ & \\
\chandra & 790 & 1999-12-27 & 44.1 & 5.0 & $1.4\pm0.4$ & $0.27\pm0.18$ &  $0.7\pm0.2$ & low state\\
         &     &             &      &     &  $28.7\pm0.9$  & $0.91\pm0.06$ & $14.6\pm0.5$ & high state\\
\chandra & 383 & 2001-08-16 & 2.2 & 3.5 & $4.6\pm1.5$ & $0.43\pm0.30$ & $1.6\pm0.5$\\
ROSAT    & 600088h-0 & 1991-12-08/10 & 3.1 & 6.0 & $3.4\pm1.1$ && $22\pm7$\\
ROSAT    & 600088h-1 & 1992-06-05/07  & 24.9 & 6.0 & $0.9\pm0.2$ & & $6\pm1$\\
ROSAT    & 600714h   & 1995-01-03/07  & 10.9 & 6.0 & $4.0\pm0.6$ & & $26\pm4$\\
ROSAT    & 600714h-1 & 1995-06-13/17 & 14.0 & 6.0 & $0.3\pm0.2$ & & $2\pm1$\\
         &           & 1995-07-05/07 & 5.9& 6.0 & $0.3\pm0.2$ & & $2\pm1$\\
ROSAT    & 601111h   & 1997-12-20/27 & 16.8 & 6.0 & $1.1\pm0.3$ & & $7\pm2$\\
\ein     & 583/2083  & 1979-07-05/08   & 27.4 & 11.5  & $1.5\pm0.3^{ *}$ & & $18\pm4$\\
\noalign{\smallskip}
\hline
\noalign{\smallskip}
\end{tabular}
\label{observations}

Notes and references:\\
$^{ *~}$: according to \citet{1984ApJ...286..491F}\\
$^{ {\dagger}~}$: The slightly higher HR during
observation 0125960101 compared to 0125960201 reflects the difference between
medium and thin filter and most likely not a change of the spectrum of \rxj.\\
$^{ **}$: 0.5--2.4 keV absorption corrected luminosity assuming an absorbed 
power law spectrum (\nh = 1.9\hcm{21}, photon index $\Gamma=1.7$) and a distance of \n253 of
2.58 Mpc \citep{1991AJ....101..456P} which we use throughout the paper\\
\end{table*}

For the detailed analysis of \rxj\ we used data from \xmm, \chandra, ROSAT HRI,
and \ein.
Table~\ref{observations} summarizes the observatory (col. 1), observation
identification (2), observation dates (3) and life time (4), 
extraction radius R$_{e}$ used for \xmm\ and \chandra\ count rates and light curves and 
spectra (5),
\rxj\ count rates (6), hardness ratios for \xmm\ EPIC PN and \chandra\ ACIS S (7), 
and luminosities in the 0.5--2.4 keV band (8), an energy band covered by all observatories. 
Special events are marked under comments (9). As hardness ratio (HR) we use the
ratio of the counts in the 1.5--5.0 keV band to the counts in the 0.5--1.5 keV
band.  

During the \xmm\ observations \citep{2001A&A...365L...1J} the EPIC PN and MOS instruments 
\citep{2001A&A...365L..18S,2001A&A...365L..27T} were operated in the full
frame mode resulting in  a time
resolution of 73.4 ms and 2.6 s, respectively.  
The medium filter was in front of the EPIC MOS1 camera in all observations, 
for MOS2, the thin filter was used during the observations in June 2000 and 
the medium filter in December. The PN camera operated with the medium filter
during  observation 0125960101 and with the thin filter for the other
observations. These filter changes do not influence the \rxj\ observations
significantly as source counts are mainly detected above 0.7 keV (see below). 
In this energy band the transmission of the thin and medium filter do not differ
significantly.
We used all EPIC instruments for the imaging and position determination of \rxj.
For the timing and spectral investigations we concentrated on the EPIC PN camera 
which gives about twice the number of photons and checked the results for 
consistency and at the beginning of the observing window using the MOS cameras. 
While the \xmm\ point spread function (PSF) normally requires extraction radii
$>30$\arcsec\ to encircle $>80\%$ of the source photons, 
we had to restrict ourselves to 10\arcsec\ to avoid
contamination from source VP99 X18 (see Fig.~\ref{x_images}).

Three \chandra\ ACIS S observations of \n253\ (see Table~\ref{observations}) 
were obtained from the \chandra\ Data Archive
(http://asc.harvard.edu/cgi-gen/cda). The instrument was operated in the full frame
mode (3.2 s time resolution). \rxj\ is positioned in the back-illuminated CCD chip S3
during observation 969 and 383, in the front-illuminated chip S2 during 790.
Due to the larger off-axis position of \rxj\ in the latter observation, 
the \chandra\ PSF  urged us to use a larger extraction radius. 

The deep space orbits of the satellites \xmm\ and \chandra\ led to
long observation times of \rxj\ without interruption. The low earth orbits of 
the \ein\ and ROSAT satellite on the other hand, led to observations split in
many short intervals of typically less than 1\,500 s. 

\n253 was observed with the ROSAT PSPC and HRI in several observing runs from 
Dec 1991 to Dec 1998. We restricted our analysis on 75.5 ks HRI observations 
as only this ROSAT detector fully resolves \rxj\ from VP99 X18.
The HRI observation 601113h (July 1998, 2 ks) could not be used for the 
investigations due to attitude problems. 

\rxj\ can not be resolved from nearby sources with the \ein\ IPC.
However, the source was clearly detected with the HRI (FT84) during a bright
state. The nearby source VP99 X18 was not detected. For additional timing 
analysis we extracted the \n253\ HRI data from the \ein\ Data Archive at HEASARC. 

The data analysis was performed using tools in the SAS v5.3.3, CIAO v2.2,
EXSAS/MIDAS 1.2/1.4, and 
FTOOLS v5.1 software packages, the imaging application DS9 v2.1, the timing 
analysis package XRONOS v5.19 and spectral analysis software XSPEC v11.2.   

For the time variability investigations all \rxj\ event times were 
corrected to solar system barycenter arrival times.

\subsection{Time variability}
\begin{figure}
   \resizebox{\hsize}{!}{\includegraphics[angle=-90]{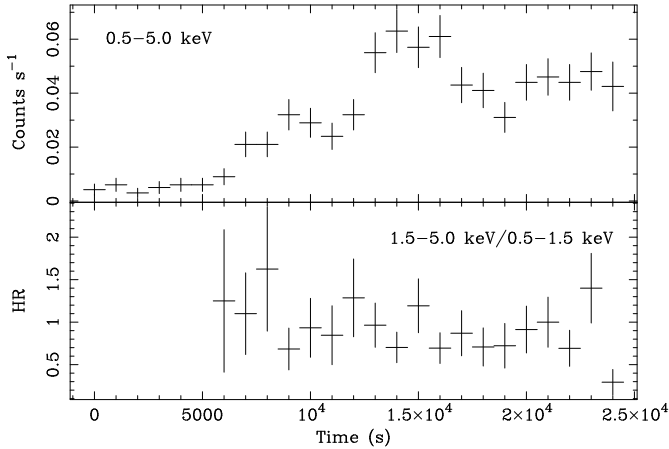}}
    \caption[]{\xmm\ EPIC PN light curve and hardness ratio of observation 
    0110900101 integrated over 
    1000 s. Time zero corresponds to MJD 51892.071251 (solar 
    system barycenter corrected). 
    \label{pn_gt2_lc}}
\end{figure}
\begin{figure}
   \resizebox{\hsize}{!}{\includegraphics[angle=-90]{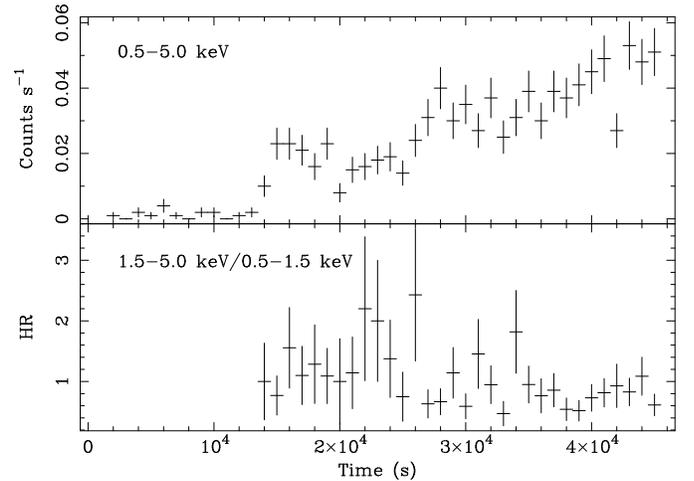}}
    \caption[]{\chandra\ ACIS S light curve and hardness ratio of  
    observation 790 integrated over 
    1000 s. Time zero  corresponds to MJD 51539.114225 (solar 
     system barycenter corrected). 
    \label{ch_790_lc}}
\end{figure}
\begin{figure*}
  \resizebox{6.cm}{!}{\includegraphics[bb=94 188 514 610,clip]{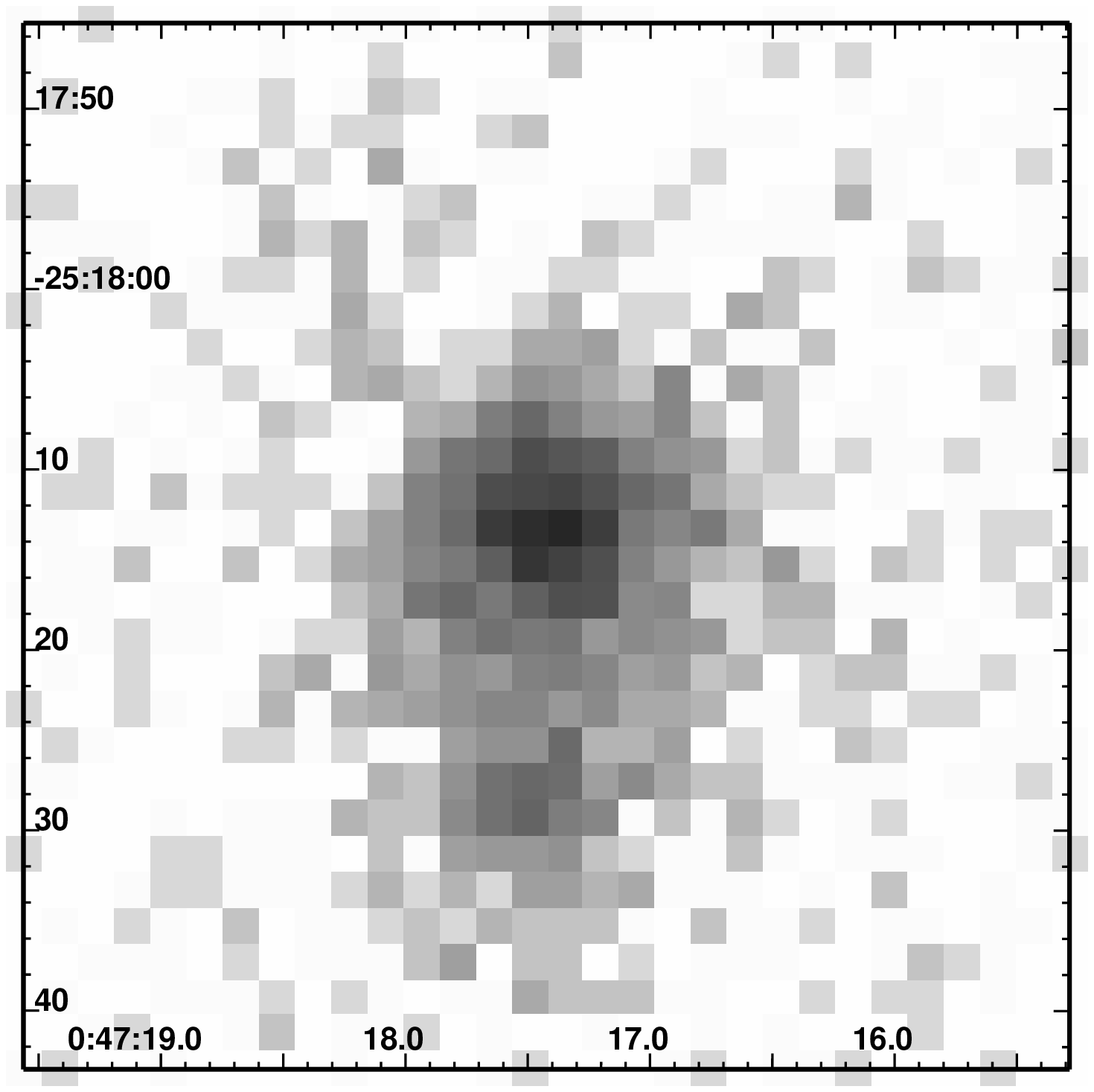}}
  \resizebox{6.cm}{!}{\includegraphics[bb=94 188 514 610,clip]{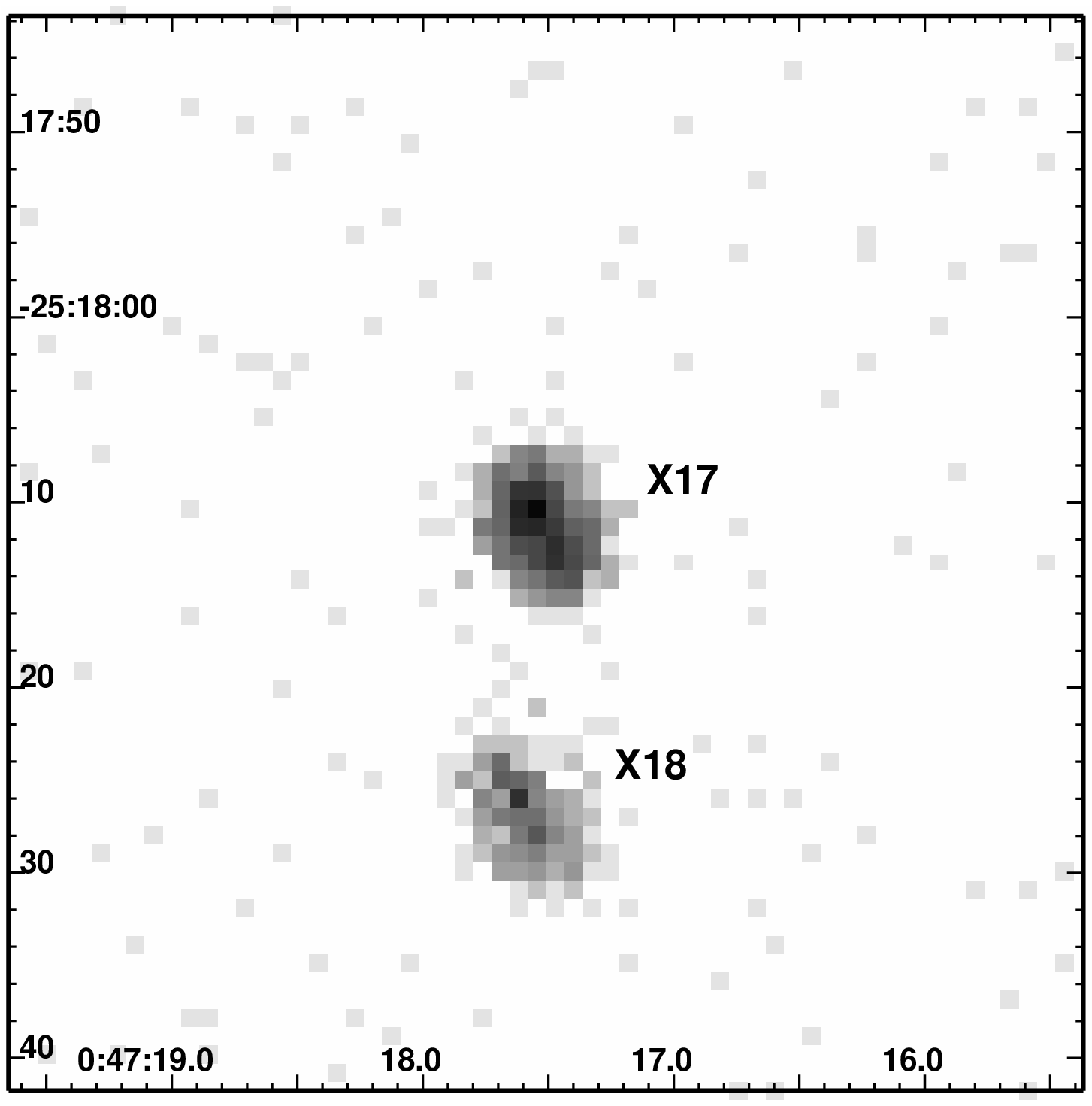}}
  \resizebox{6.cm}{!}{\includegraphics[bb=94 188 514 610,clip]{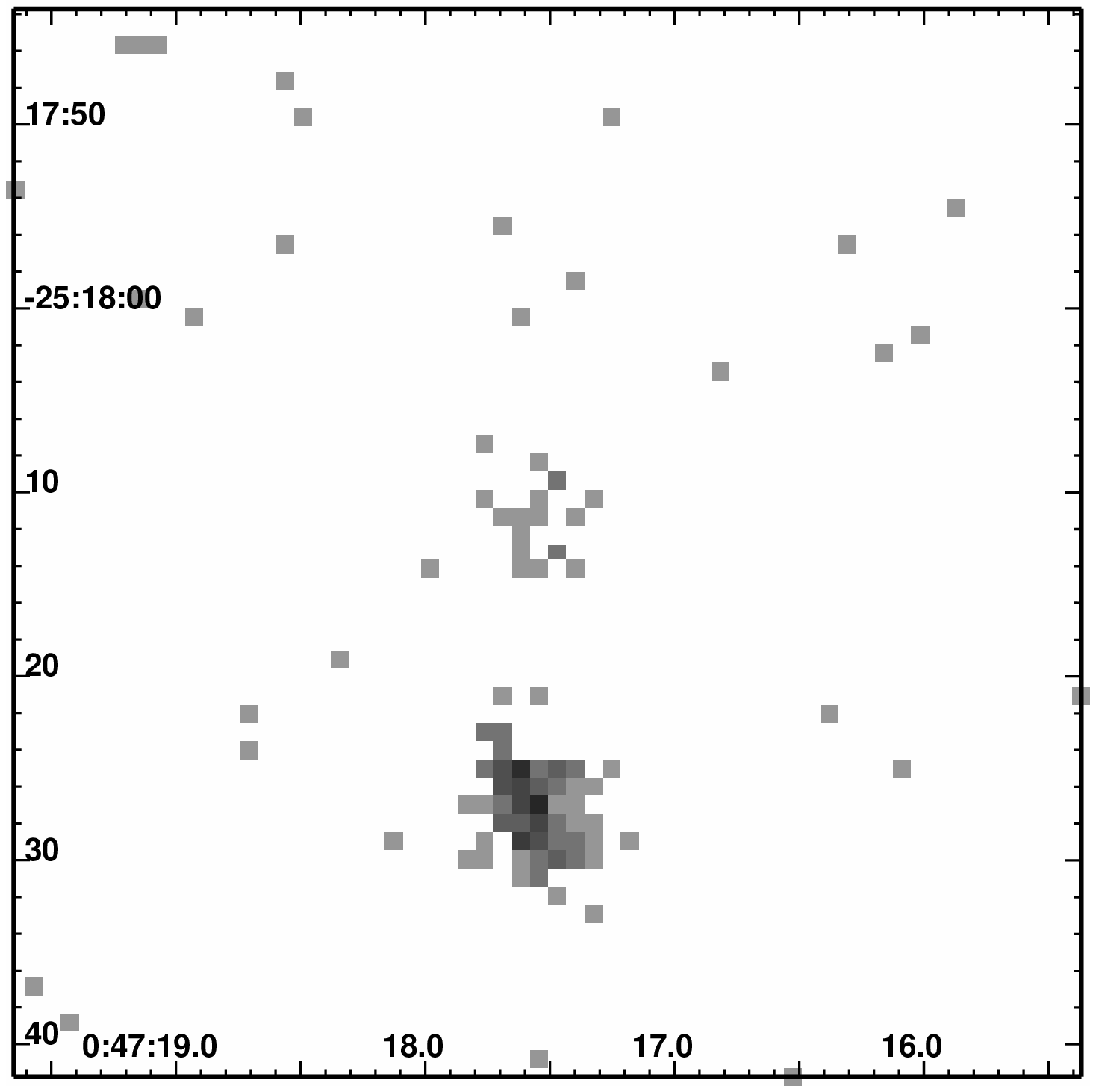}}
    \caption[]{
  Logarithmically-scaled, grey-scale images of the \rxj\ region of the 
  \xmm\ EPIC high state in observation 0110900101 and
  \chandra\ ACIS S high and low state in observation 790 (from left to
  right).
  The images (RA, Dec J2000.0) were binned with a pixel size of 2\arcsec,
  1\arcsec, 1\arcsec\  
  in the 0.5--5.0 keV band for times after 8\,500 s in Fig.~\ref{pn_gt2_lc}, 
  after 14\,500 s and before 13\,500 s in Fig.~\ref{ch_790_lc}, and 
  have maxima of 66, 92, and 9 counts per pixel, respectively. The sources X17 
  (\rxj) and X18 are labeled in the center image.
  
    }
    \label{x_images}
\end{figure*}
The EPIC PN as well as the MOS light curves for \xmm\ observation 0110900101 
show egresses from low to high state at
solar system barycenter corrected MJD $51892.146\pm0.006$ (
for EPIC PN see Fig.~\ref{pn_gt2_lc})
with residual emission during low state which partly may be contaminated from 
emission collected from the PSF wing of VP99 X18. During the high state the
average HR was 0.86 and constant within the errors (average HR 0.86, see 
Table~\ref{observations}), the HR during the low state (0.5) indicates a softer
spectrum and/or less absorption.  
During each of the \xmm\ observations 0125960101 and 0125960201 the flux was 
constant within the errors (see Table~\ref{observations}) 
and the source was a factor of 1.9 
brighter than in the 0110900101 low state, a factor of more than five below that
in the 0110900101 high state. The HRs during these observations 
on the other hand, appear to be quite similar to the  0110900101 low state.

The light curve for the \chandra\ observation 790 shows an egress from low to high
state at solar system barycenter corrected MJD $51539.276\pm0.006$ 
(Fig.~\ref{ch_790_lc}). During the high state the
HR defined in the same energy bands as for \xmm\ EPIC, was 0.91 and again 
constant within the errors. The residual emission during the low state clearly
originates from \rxj\ as is demonstrated in the right part of 
Fig.~\ref{x_images} and corresponds to a count rate of 1.4 ct ks$^{-1}$.  
The HR again is significantly smaller than during the high state. 
The low state flux was about a factor of 2 below that in the \xmm\ low state.
During \chandra\ observation 969, \rxj\ flux and HR were constant 
within the errors and the source about a factor of 1.4 brighter with similar
average HR than that measured in the high state of 790.
During \chandra\ observation 383 the  \rxj\ was a factor of 12 below that during
969, but a factor of 2.3 higher than in the 790 low state, 
and showed a small HR (see Table~\ref{observations}). 
However, one has to keep in mind that the \chandra\ ACIS S count rates and HRs 
of \rxj\ can not
directly be compared to the others as they originate from a CCD with a 
different energy response function. The same is even more true if one wants to
compare \chandra\ and \xmm\ count rates and HRs.

Figure~\ref{x_images} shows images (1\arcmin\ to a side) of the \rxj\ area in 
the 0.5--5.0 keV band. For \xmm\ all EPIC cameras are combined for the high
state of observation 0110900101. \rxj\ is the source to the north, VP99 
X18 the one to the south.
The \chandra\ ACIS S images of observation 790 are divided in a low and high
state image. \rxj\ is clearly detected also during the low state.

We searched for pulsations within the bright parts of the \xmm\ and 
\chandra\ observations in the frequency range 10$^{-3}$--3 Hz. The strongest
signal was found at 29.5 s in the \xmm\ observation 0110900101 which, following
the Rayleigh Z$^2_{\rm n}$ method \citep{1983A&A...128..245B} as described in
\citet{2002A&A...391..571H} yields only a 71\% confidence level ($\sim 1
\sigma$).

\begin{figure}
   \resizebox{\hsize}{!}{\includegraphics[]{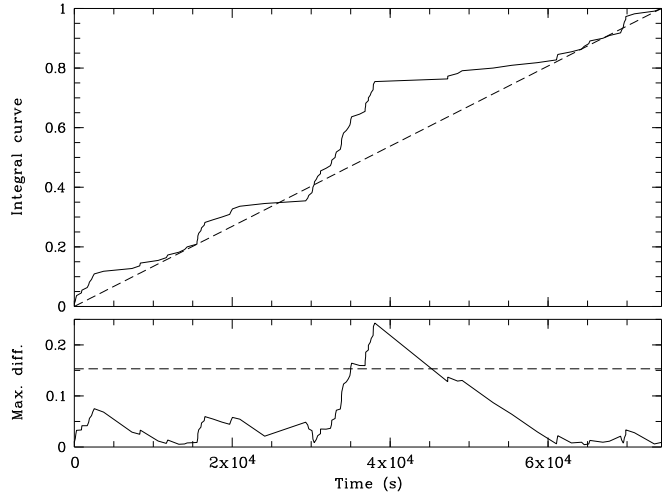}}
    \caption[]{ROSAT HRI integrated light curve of \rxj\ events (selected 
    in a circle around the source with 6\arcsec\ radius) 
    over the observation time (top). The dashed line shows the 
    expected light curve for a constant source of the same luminosity. 
    The difference between the light curve of the constant source and 
    the observed counts is given below. A 99\% variability threshold according to a
    Kolmogorov-Smirnov test is shown as dashed line. 
    \label{hri_ks}}
\end{figure}
During the ROSAT HRI observations covering 6 years, 110 photons were detected
within 6\arcsec\ from \rxj\ in an effective observing time of 75.5 ks. 
Due to the low ROSAT HRI background 
less than 10 of these counts may be due to field background. The source photons 
were not distributed evenly over the observing time but clearly showed strong 
intensity changes and even short term flares as is demonstrated by a
Kolmogorov-Smirnov test against constancy (see Fig.~\ref{hri_ks}). 

\subsection{Improved position}
\begin{figure}
   \resizebox{\hsize}{!}{\includegraphics[clip]{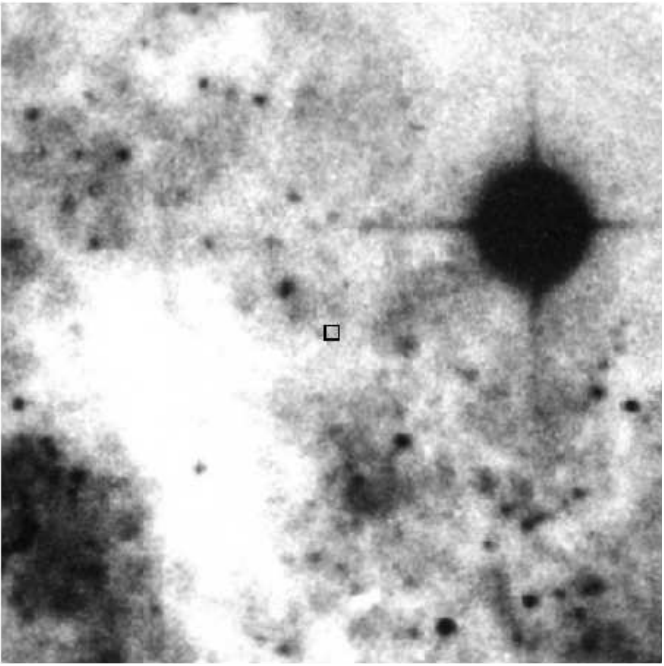}}
    \caption[]{Position of \rxj\ is shown on an extract of a deep blue \n253\ image
    \citep{1994AJ....108.2102S} covering the same area as Fig.~\ref{x_images} and
    was kindly provided by Akiko Kawamura. The square 
    (1\arcsec\ at a side) marks the improved position. The image was aligned 
    using USNO-B positions in the field \citep{monet03}. The residual error 
    of the solution is about 1\arcsec\ .    
    \label{x17_opt}}
\end{figure}
VP99 give the position of \rxj\ with an error of 2\farcs7 which is mostly
determined by the included systematic error of 2\farcs5. The much bigger  
number of photons detected with the \xmm\ EPIC and \chandra\ ACIS S detectors
allows us to determine a significantly improved source position. Systematic
errors can be reduced by adjusting the positions using the well determined
optical \citep[0\farcs2 error using USNO-B positions, ][]{monet03} and 
X-ray positions of several QSOs in the field of view. 

For \chandra\ ACIS S observation 790 the VP99 source X22 identified with a QSO 
(optical blue magnitude of 19.9, redshift 1.25) is detected in chip S3 close to
the center of the field of view. During observation 969 it is positioned at a larger off-axis 
angle in chip S4.
For \xmm\ EPIC besides VP99 X22 the sources X4 and X58 are clearly detected (QSOs with 
optical blue magnitudes of 20.3 and 18.5 and redshifts of 1.02 and 0.66,
respectively). Averaging the resulting positions from \chandra\ observations 
790 and 969 as well as from the individual EPIC instruments in observation  
0110900101 we get a highly improved position for \rxj: 
$\alpha$=00$^{\rm h}$47$^{\rm m}$17\fs51, $\delta$=$-$25\degr18\arcmin11\farcs2 
(J2000), with an remaining error radius of 0\farcs3 mainly determined by
systematics. See Fig.~\ref{x17_opt} for an overlay of the improved position on a
deep optical blue image.

\subsection{Energy spectra}
\begin{table*}
\caption{Spectral modeling results for \rxj. Only \chandra\ ACIS S and \xmm\ EPIC PN times
in high state were used.
In the case of $\chi^2/\nu\le2.0$, 90$\%$ errors are given}
\label{spectralfits}
\begin{flushleft}
\begin{tabular}{lrrrlrrrrr}
\hline
\noalign{\smallskip}
Observation & $T_{\rm int}$ & Raw count rate & DOF & Model$^*$ & $N_{\rm H}$ & 
     $\Gamma$&$T$  & $L_{\rm X}^{**}$ & $\chi^2/\nu$\\
& (ks) &($10^{-2}$ ct s$^{-1}$)& & & ($10^{21}$~cm$^{-2}$) & & (keV) &
    ($10^{38}$ erg s$^{-1}$) \\
\noalign{\smallskip}
\hline
\noalign{\smallskip}
\xmm & 16.5 & 3.8$\pm$0.7 & 17 & POWL  & $2.3_{-0.7}^{+1.0}$ &   
$1.9_{-0.3}^{+0.2}$&&  $2.6_{-0.6}^{+0.7}$  & 1.63\\ 
\noalign{\smallskip}   
EPIC PN &&&& BREMS & $1.6_{-0.6}^{+0.8}$ && $6_{-2}^{+4}$       
&  $2.3_{-0.4}^{+0.4}$ &1.77  \\\noalign{\smallskip} 
0110900101&&&&THPL  &   12         &&   1.0        & 5.8  & 7.1 \\ 
\noalign{\smallskip}\hline
\noalign{\smallskip}
\chandra & 14.2  & 5.7$\pm$0.3 & 21 & POWL  & $1.9_{-0.8}^{+0.8}$ &  
$1.7_{-0.2}^{+0.2}$&&  $4.1_{-0.7}^{+1.1}$  &0.69\\
\noalign{\smallskip}
969&&&& BREMS & $1.2_{-0.6}^{+0.7}$ && $8_{-3}^{+6}$      
&  $3.9_{-0.4}^{+0.4}$ & 0.69 \\
\noalign{\smallskip}
&&&&THPL  &      10      && 1.0          & 6.9  & 9.7 \\
\noalign{\smallskip}\hline
\noalign{\smallskip}
\chandra & 31.0 & 3.2$\pm$0.1 &26  & POWL  & $1.8_{-0.9}^{+1.0}$  
& $1.8_{-0.2}^{+0.2}$& & $3.0_{-0.7}^{+0.7}$  &1.10\\
\noalign{\smallskip}
790&&&& BREMS & $1.0_{-0.7}^{+0.7}$ && $8_{-3}^{+5}$      
&  $2.7_{-0.3}^{+0.3}$ & 1.11 \\\noalign{\smallskip}
&&&&THPL  &    9.0        &&     1.0      & 4.5  &6.3  \\
\noalign{\smallskip}\hline
\noalign{\smallskip}
\end{tabular} 
\end{flushleft}
$^*$\ : \hskip.3cm THPL = thin thermal Plasma with solar abundance, 
BREMS = thermal bremsstrahlung, POWL = power law\par
$^{**}$: \hskip.3cm In the 0.5--7.5 keV band, corrected for total absorption,
corrected for extraction radii and vignetting\par
\end{table*}   
As discussed in Sect. 2.1, HR changes of \rxj\ in different brightness regimes 
indicate spectral variability which may be either caused by reduced absorption
during the low intensity states or due to changes of the stiffness of the source
spectrum. Unfortunately, the source count rates do not allow a more detailed
investigation. For the brighter states of \rxj, that is
during \chandra\ observation 969 as well as during the high state of
\chandra\ observation 790, and \xmm\ observation 0110900101 
more than 600 counts were
collected from the source which allowed us a rough spectral analysis.
\rxj\ photons were extracted using the same areas as for the time variability analysis
and spectra were grouped in channels with at least 30 source plus background photons.  
Due to the limited number of photons 
we just fitted absorbed one component spectral models: power law (POWL), 
bremsstrahlung (BREMS), 
and thin thermal plasma with element abundance fixed to solar (Raymond-Smith, 
THPL).
The resulting integration times, raw count rates and degrees of freedom (DOF) 
are listed in Table~\ref{spectralfits} together with the results of the fits. 

The absorbed POWL model fits show the lowest $\chi^2/\nu$ values. Absorbing column and 
photon index $\Gamma$ of all observations coincide within the errors. However,
only for the \chandra\ observations the fit is formally acceptable. 
Several factors may explain the unacceptably high $\chi^2/\nu$ of the POWL 
fit for the \xmm\ EPIC PN spectrum of \rxj. To the one hand, there may be 
contributions to the spectrum from the nearby source X18. On the other
hand, to get sufficient statistics for the fit we accepted all photons in
the extraction region, not rejecting events on or close to a CCD column
with slightly increased offset which passes through the extraction region.
The BREMS spectra exhibit marginally higher
$\chi^2/\nu$ values than the POWL fits and find
high temperatures (around 7 to 10 keV). 
The absorbing column needed in addition to the POWL and BREMS models 
is $\sim2\times10^{21}$~cm$^{-2}$, significantly above the absorption expected
within the Milky Way in this direction 
\citep[$1.3\times10^{20}$~cm$^{-2}$,][]{1990ARA&A..28..215D}.
Neither the \chandra\ nor the \xmm\ data can be described by 
a THPL spectrum.
  
The best fit POWL model parameters of \chandra\ observation 969 are used 
in Table~\ref{observations} for conversion from count rates to intrinsic source
luminosities for observations of all observatories. We are aware that the 
smaller HR values during low intensity states indicate a higher low energy flux
and therefore a different spectrum. This may lead to an overestimation of the
intrinsic low state luminosities in the table.

\section{Orbital period determination}
The intensity changes of \rxj\ from low to high in the \xmm\ 
observation 0110900101 and in the \chandra\ observation 790 are not resolved 
by the 1000 s time resolution of our plots.  Shorter time bins do not help as
the source is not bright enough. 
In the following we interpret the intensity
changes as egresses from eclipse of an XRB. 
The times of egresses have to be 
separated by an integer number of periods and therefore give a first constrain on 
the possible orbital period p of \rxj:

p = ($352.870\pm0.012$) d / n ; n = 1,2,3,...              \hfill  (1)

without exactly knowing the number n of periods between the two egresses. 
The relative period error is 3.3\expo{-5}. 

The \chandra\ observation 790 is the longest continuous observation of \rxj\ 
covering 0.5092 d, however only part of an orbit. This together with Eq. 1 
constrains the number of possible periods to 692 (covering the range 0.5099 d 
and 352.870 d).

During \chandra\ observation 969    
the flux of \rxj\ was constant within the errors and even brighter 
than during the 790 bright state. The source therefore undoubtedly was out of 
eclipse. If we use the conditions that the non-eclipse observations are not 
allowed to include orbit
phases covering the minimum eclipse duration (0.14 d during the \chandra\
observation) and assuming that the eclipse phase is  
longer than 0.13. This constrains the number of possible periods to 457 
(between 0.5129 d and 70.574 d).

We can try to further reduce the number of acceptable periods with the help of 
the other \xmm\ and \chandra\ observations and in addition use the sparse ROSAT 
and \ein\ HRI photons. This selection however has to be taken with care
as it is hampered by the limited knowledge of the luminosity state of
\rxj\ during these observations and the limited photon statistics. 

\xmm\ observations 0125960101 (duration 0.451 d) and 0125960201 (0.164 d)  
were separated by 0.174 d. \rxj\ was again 
constant within the errors during each observation and also from observation
to observation, however only a factor of 1.9 brighter than during 
eclipse of 0110900101 with comparable hardness ratio values. As the residual 
flux during eclipse may vary from 
eclipse to eclipse similar to the source flux out of eclipse we can not decide
if the observations are out of eclipse (I) or if one eclipse covers both 
observations (II). The condition (I) reduces the number of allowed periods to
141 (between 0.6416 d and 70.574 d). If we assume for condition (II) 
a maximum eclipse duration
of 0.25 in phase (similar to that reported for Cen X-3), the period has 
to be longer than 3.156 d, the two observations at the right phase and 
\chandra\ 969 not within the same phase interval. The condition (II) reduces 
the number of allowed periods to 10 (between 4.1031 d and 32.079 d).
 
Also for 
\chandra\ observations 383 we can not decide if the source was in or out
of eclipse as the source is only a factor of two brighter than during eclipse 
of 790 and shows similar hardness ratio. 
For 23 of the 141 period candidates of condition (I) and 3 of the 10 for (II) 
the source would be in eclipse during \chandra\ observation 383.

The detection of two eclipse egresses within the few \xmm\ and \chandra\
observations of less than 0.5 d duration strongly favors a short orbital period
of \rxj. If we constrain the period to less than 10 d there only remain 118 periods for
condition (I) and 5 for (II) to be
investigated. During 18 (1)  of those \rxj\ would be in eclipse during \chandra\ 
observation 383.

To further reduce the number of acceptable periods we folded the ROSAT HRI
counts detected from the \rxj\ area over the period ranges allowed by the 
candidate periods. We checked if the source could have remained in low 
intensity (eclipse) for at least the duration of the measured longest low 
states (0.14 d) or for 0.13 of the candidate period and 
if this low intensity time was compatible in phase with the \chandra\ eclipse 
egress for period candidates derived with condition (I). For condition (II)
the eclipse had to last as long as indicated by the \xmm\ observations. 
Many of the candidate periods could be rejected using these criteria. 
Some periods can not be rejected because the expected eclipse phase is not 
covered by ROSAT exposure. Some periods however show only background counts 
at the expected eclipse phases even when the ROSAT exposure is high. 
All periods selected under condition (II) are not acceptable using these
criteria. Under condition (I), we get 14 candidate periods with a lower error
range. During three of the periods, \rxj\ is in eclipse during \chandra\ 
observation 383. 

In a last step, 
the candidate periods were checked for consistency with the \ein\ 
observations (collected during a \rxj\ high state in July 1979, 
see Table~\ref{observations}). Only seven periods, two with \chandra\ 
observation 383 in eclipse, pass this test (see Table~\ref{periods}). 
The periods above 5 days are less probable as the chance to detect two eclipse
egresses during the \xmm\ and \chandra\ observations would have been low. From
the remaining periods the ones with allowed eclipse duration longer than 0.13
in phase seem to be more probable. 

Taking all the selection criteria into account
1.470243 d seems to be the best period candidate followed by 3.207928~d.
Additional observations of \rxj\ in the high state are urgently needed to
finally decide on the correct orbital period.
\begin{table}
\caption[]{Allowed orbital periods of \rxj.}
\begin{tabular}{lrrrr}
\hline\noalign{\smallskip}
\multicolumn{1}{c}{Period} &
\multicolumn{1}{c}{Error$^{*}$} & 
\multicolumn{1}{c}{Ecl. dur.} &
\multicolumn{1}{c}{Obs. 383} &
\multicolumn{1}{c}{Comment} \\ 
\noalign{\smallskip}
(d)& (\oexpo{-6} d)& ($^{**}$)&($^{***}$)&($^{****}$)\\
\noalign{\smallskip}\hline\noalign{\smallskip}
1.470243& 10 & 0.15 && B\\
2.484902& 10 & 0.13 && B \\
2.778391& 10 & 0.13 &E& B\\
3.207928& 10 & 0.14 && \\
4.969849& 20 & 0.13 &E& B\\
6.190937& 20 & 0.14 && B, E-\\
7.671308& 20 & 0.15 && E-\\
\noalign{\smallskip}
\hline
\noalign{\smallskip}
\end{tabular}
\label{periods}

Notes:\\
$^{ *~~~}$: determined to achieve longest possible eclipse duration\\
$^{ **~~}$: maximum allowed eclipse duration ($\Delta$phase)\\
$^{ ***~}$: E if \rxj\ in eclipse during \chandra\ 383\\
$^{ ****}$: B period at boundary of allowed window, E- no \ein\ exposure during
eclipse\\
\end{table}

\section{Discussion}
In the last sections we have shown that the variety of observations of \rxj\ can be
explained in an eclipsing XRB scenario. \rxj\ is the first eclipsing XRB
identified outside the Local Group. 

The length of eclipse and the 
luminosity during its high state of $\sim2\times10^{38}$ erg s$^{-1}$, i.e. at
the Eddington limit of a 1.4 M$_{\sun}$\ neutron star,
are most easily explained by a HMXB with a neutron star as companion. 
Similar luminosities were reported for the eclipsing HMXBs SMC X-1 and 
LMC X-4 during the high state.  Detection of pulsations in the tens of second 
range or shorter are expected in such systems. For a faint source like \rxj\ 
such short pulsations are difficult to detect as 
time delays of photon arrival times due to the motion in the binary orbit 
will smear out the periodic signal. 

Accreting XRBs normally show pulsation averaged X-ray
energy spectra above 1 keV that can be modeled by power laws
with high energy cut-offs. The photon indices of the power law spectra range
between 0.8 and 1.5 for most pulsars, but some sources show rather soft spectra
\citep[see the review of ][]{1989PASJ...41....1N}. The steep
power law slope of 1.9 in the brighter states puts \rxj\ in this special
category.  

For most eclipsing XRB residual emission during eclipse was measured. The
emission can be explained by re-processing of primary X-rays originating from the 
bright area at the magnetic poles of the neutron star, in an extended accretion
disk corona (which is not fully occulted) or by scattering in the companion 
atmosphere/stellar wind. Residual emission of up to $\sim$10\% was reported 
\citep{1996PASJ...48..425E,1991A&A...252..272H} depending on system geometry and
wind density. The
\rxj\ residual emission of  $\sim$5\% is well within these limits.

As discussed in Sect. 3,~\rxj\ was most likely out of eclipse during 
the \xmm\ observations in December 2000 and the \chandra\ observation in August
2001. During these observations the source luminosity was up to a factor of ten 
lower than during the high states (luminosities close to Eddington) 
indicating clear long term time variability. During the \ein\ observations in 
July 1979 and also during the ROSAT observations in January 1995 the source was
in high state, while during the rest of the ROSAT observations the intensity was
significantly lower (see Fig.~\ref{hri_ks}). Similar long term variability was
detected in several other low and high mass XRBs where it even may manifest
itself in a long term superorbital period (e.g. Her X-1, LMC X-4, SMC X-1)
typically on the scale of tens of days. Due to the up to now sparse time 
coverage of \rxj\ observations, such long term periodicities were not
detected, but at present also can not be ruled out. The Her X-1 and LMC X-4
long term periodicities correspond to $\sim21$ orbital periods. If a \rxj\ long
term variability follows a similar pattern, the most probable orbital 
period of 1.47 d would lead to $\sim31$ d as period of the long term variability.

During some ROSAT observation intervals of typically 1000 s duration 
there are some indications of short term flaring (see Fig.~\ref{hri_ks}). 
Similar flares up to now were not detected during \xmm\ and \chandra\ 
observations. Short term flaring was detected in several HMXBs 
\citep[e.g. LMC X-4, ][]{1985xray.symp...23P}.

The most promising orbital periods of \rxj\ do not allow us to decide if
the optical companion is a high or low mass star. LMC X-4 and SMC X-1 are HMXB 
systems with orbital periods of 1.4 d and 3.9 d, Her X-1 with an intermediate
period (1.7 d) on the other
hand is a LMXB. The
X-ray luminosity of \rxj\ in the bright state at or above the Eddington limit
and possible flaring (see above) may indicate a HMXB as the LMXB Her X-1 only
reaches about one third of the Eddington luminosity 
\citep{1972ApJ...177.L103F,1983MNRAS.202..347H} and does not show flares.

If we therefore assume that \rxj\ is a HMXB we can use the optical magnitudes of the known
systems in the Magellanic Clouds to determine the optical brightness in \n253.
The optical brightnesses of LMC X-4 and SMC X-1 
-- V of 14.0 mag and 13.3 mag  \citep{2000A&AS..147...25L} 
and distances of 50 kpc and 59 kpc, respectively 
\citep[see review by][]{1999IAUS..190..569V} -- 
correspond to (21.4--22.6) mag at the assumed distance of 2.58 Mpc to \n253. If
this distance has to be corrected to 3.94 MPc as recently indicated by 
\citet{Kara03}, not only the estimated optical magnitudes would have to be
increased by 0.9 mag but also the derived X-ray luminosities for \rxj\ would
rise by a factor of 2.3. 
The expected optical magnitude may have to be increased by up to
one optical magnitude if the absorption measured in the X-ray spectrum in the
high state (\nh=1.9\hcm{21}, see above) is due to extinction within \n253\ and 
not due to absorption in the near surrounding of the X-ray emitting area within
the binary system  \citep{1995A&A...293..889P}. Additional extinction within 
\n253\ may well be present as \rxj\ seems to be positioned at the edge of a
dark cloud (see Fig.~\ref{x17_opt}). A preliminary analysis of the image 
indicates that stars of visual magnitude 23 should just be visible. There is
a faint object at the east edge of the error box which could be the optical
counterpart. However, if \rxj\ is a LMXB similar to Her X-1 
\citep[14 mag at a distance of 6 kpc, see e.g.][]{1983MNRAS.202..347H}, 
it would be fainter than 27 mag at the distance of 
\n253, very difficult to be separated from the many other stars of similar
brightness in the \n253\ surrounding, and not be detectable in current optical 
imaging. 
Only deeper time resolved imaging and spectroscopy can lead to a secure
identification of \rxj as a HMXB and solve the high mass -- low mass ambiguity. 

\section{Conclusions and outlook}
Starting from an \xmm\ observation in which \rxj\ changed state from low to
high, we found a similar transition in a \chandra\ observation. 
Interpreted as eclipse
egresses we determined candidate orbital periods for the source. With the help
of archival \xmm, \chandra, ROSAT and \ein\ observations we selected as most
probable orbital periods 1.47029~d and 3.20793~d. \rxj\ is one of the few
eclipsing X-ray binary systems and the first detected outside the Local Group.
No pulsation period was found. The source is clearly variable on time scales
of tens of days and also short time flaring (1000 s time scale) is indicated. 
The energy spectrum of the source during the high state can be described by a
hard power law spectrum with absorption significantly higher than the foreground
absorption in the Milky Way. Using the improved source position we indicate
a possible optical identification. 
\rxj\ is most likely a HMXB system similar to SMC X-1 and LMC X-4. 

Two long observations of \n253\ accepted for the \xmm\  and
\chandra\ program (140 ks and 85 ks, respectively) 
hopefully will detect \rxj\ in the high state and allow
us to finally decide on the orbital period of the source and enable sensitive
neutron star rotation searches from seconds to tens of seconds. 
The \xmm\ observation would cover one full binary orbit if the 1.5 d period is
correct and finally clarify the correct orbital solution. 
A search for sub-second periods will only be possible in dedicated
observations of \rxj\ with higher time resolution (that could not fully cover 
the galaxy).

\begin{acknowledgements}
    The \xmm\ project is supported by the Bundesministerium f\"{u}r
    Bildung und Forschung / Deutsches Zentrum f\"{u}r Luft- und Raumfahrt 
    (BMBF/DLR), the Max-Planck Society and the Heidenhain-Stiftung.
\end{acknowledgements}
\bibliographystyle{apj}
\bibliography{./3430,/home/wnp/data1/papers/my1990,/home/wnp/data1/papers/my2000,/home/wnp/data1/papers/my2001}

\end{document}